\documentclass{ws-procs9x6}

\usepackage{graphicx,amssymb}

\def\bea{\begin{eqnarray}}
\def\eea{\end{eqnarray}}
\def\be{\begin{equation}}
\def\ee{\end{equation}}
\newcommand{\ub}[1]{\underline{#1}}

\def\del{\partial}
\def\psibar{\overline{\psi}}
\def\bra{\langle}
\def\ket{\rangle}
\def\tr{{\rm tr}}

\def\ep{\epsilon}
\def\g{\gamma}

\begin{document}

\title{%
Nonperturbative solution of Yukawa theory and gauge theories%
\footnote{\uppercase{P}reprint \uppercase{UMN-D-04-4},                         
to appear in the proceedings of the sixth workshop on 
\uppercase{C}ontinuous \uppercase{A}dvances in \uppercase{QCD},
\uppercase{M}inneapolis, \uppercase{M}innesota, 
\uppercase{M}ay 13-16, 2004.}
}

\author{John R. Hiller}

\address{Department of Physics \\
University of Minnesota-Duluth \\
Duluth, MN 55812 USA \\
E-mail: jhiller@d.umn.edu}

\maketitle

\abstracts{%
Recent progress in the nonperturbative solution of 
(3+1)-dimensional Yukawa theory and quantum electrodynamics (QED)
and (1+1)-dimensional super Yang--Mills (SYM) theory will be
summarized.  The work on Yukawa theory has been extended
to include two-boson contributions to the dressed fermion
state and has inspired similar work on QED, where Feynman
gauge has been found surprisingly convenient.  In both cases,
the theories are regulated in the ultraviolet by the inclusion
of Pauli--Villars particles.  For SYM theory, new high-resolution
calculations of spectra have been used to obtain thermodynamic
functions and improved results for a stress-energy correlator.
}

\section{Introduction}

Numerical techniques can be successfully applied to the nonperturbative 
solution of field theories quantized on the light 
cone.\cite{Dirac,PauliBrodsky,DLCQreview}\  Unlike lattice
gauge theory,\cite{lattice}\ wave functions are computed directly in a
Hamiltonian formulation.  The properties of an eigenstate can then be 
computed relatively easily.  There have been a number of successes in
two-dimensional theories,\cite{DLCQreview}\ but in three or four
dimensions the added difficulty of regulating and renormalizing
the theory has until recently limited the success of the approach.

Here we discuss recent progress with two different yet related
approaches to regularization.  One is the use of Pauli--Villars (PV)
regularization\cite{PV,bhm-previous,OneBoson,bhm,qed}\ and the 
other supersymmetry.\cite{SDLCQreview}\  The particular applications 
to be discussed are to Yukawa theory 
and QED in 3+1 dimensions with PV fields
and to super Yang--Mills (SYM) theory in 1+1
dimensions.  In the latter case, extension to 2+1 dimensions has
already been done;\cite{3Dsdlcq}\ however, the most recent 
developments have used two dimensions as a proving ground.  
There we consider in particular a stress-energy 
correlator\cite{N88correlator,correlator,N22}\ and analysis of 
finite-temperature effects.\cite{FiniteTemp}
  
The light-cone coordinates\cite{Dirac}\ that we use are defined by
$x^\pm = x^0\pm x^3$, $\vec{x}_\perp=(x^1,x^2)$,
with the expression for $x^\pm$ divided by $\sqrt{2}$ in the
case of supersymmetric theories.  Light-cone three-vectors
are denoted by an underline: $\ub{p}=(p^+,\vec{p}_\perp)$.

The key elements of the PV approach are the introduction of
negative metric PV fields to the Lagrangian, with couplings
only to null combinations of PV and physical fields; the use
of transverse polar coordinates in the Hamiltonian eigenvalue
problem; and the introduction of special discretization of
this eigenvalue problem rather than the traditional
momentum grid with equal spacings used in discrete light-cone
quantization.\cite{PauliBrodsky,DLCQreview}\  The choice of
null combinations for the interactions eliminates instantaneous
fermion terms from the Hamiltonian and, in the case of QED,
permits the use of Feynman gauge without inversion of a covariant
derivative.  The transverse polar coordinates allow use of
eigenstates of $J_z$ and explicit factorization from the
wave function of the dependence on the polar angle; this
reduces the effective space dimension and the size of the
numerical calculation.  The special discretization allows
the capture of rapidly varying integrands in the product
of the Hamiltonian and the wave function, which occur for
large PV masses.

For supersymmetric theories, the technique used is supersymmetric
discrete light-cone quantization (SDLCQ),\cite{Sakai,SDLCQreview}\ 
which is applicable to theories with enough supersymmetry to
be finite.  This method uses the traditional DLCQ grid
in a way that maintains the supersymmetry exactly within the 
numerical approximation.  The symmetry is retained by discretizing
the supercharge $Q^-$ and computing the discrete Hamiltonian
$P^-$ from the superalgebra anticommutator $\{Q^-,Q^-\}=2\sqrt{2}P^-$.
To limit the size of the numerical calculation, we work in the
large-$N_c$ approximation; however, this is not a fundamental
limitation of the method.


\section{Yukawa theory}

The Yukawa action with a PV scalar and a PV fermion is
\bea
S&=&\int d^4x
\left[\frac{1}{2}(\del_\mu\phi_0)^2-\frac{1}{2}\mu_0^2\phi_0^2
-\frac{1}{2}(\del_\mu\phi_1)^2+\frac{1}{2}\mu_1^2\phi_1^2\right.  \\
 &&+\frac{i}{2}\left(\psibar_0\g^\mu\del_\mu-(\del_\mu\psibar_0)\g^\mu\right)
     \psi_0
  -m_0\psibar_0\psi_0  \nonumber \\
&&
  -\frac{i}{2}\left(\psibar_1\g^\mu\del_\mu-(\del_\mu\psibar_1)\g^\mu\right)
      \psi_1+m_1\psibar_1\psi_1 \nonumber \\
&&\left.
      -g(\phi_0 + \phi_1)(\psibar_0 + \psibar_1)(\psi_0 + \psi_1)\right].
\nonumber
\eea
From this we obtain the light-cone Hamiltonian
\bea \label{eq:YukawaP-}
\lefteqn{P^-=
   \sum_{i,s}\int d\ub{p}
      \frac{m_i^2+\vec{p}_\perp^2}{p^+}(-1)^i
          b_{i,s}^\dagger(\ub{p}) b_{i,s}(\ub{p})} \\
   && +\sum_{j}\int d\ub{q}
          \frac{\mu_j^2+\vec{q}_\perp^2}{q^+}(-1)^j
              a_j^\dagger(\ub{q}) a_j(\ub{q})  \nonumber \\
   && +\sum_{i,j,k,s}\int d\ub{p} d\ub{q}\left\{
      \left[ V_{-2s}^*(\ub{p},\ub{q})
             +V_{2s}(\ub{p}+\ub{q},\ub{q})\right]
                 b_{j,s}^\dagger(\ub{p})
                  a_k^\dagger(\ub{q})
                   b_{i,-s}(\ub{p}+\ub{q})\right. \nonumber \\
      &&\left.\rule{0.5in}{0in}
           +\left[U_j(\ub{p},\ub{q})
                    +U_i(\ub{p}+\ub{q},\ub{q})\right]
               b_{j,s}^\dagger(\ub{p})
                a_k^\dagger(\ub{q})b_{i,s}(\ub{p}+\ub{q})
                    + h.c.\right\},  \nonumber
\eea
where antifermion terms have been dropped.
No instantaneous fermion terms appear because they are
individually independent of the fermion mass and cancel
between instantaneous physical and PV fermions.
The vertex functions are
\be
U_j(\ub{p},\ub{q})
   \equiv \frac{g}{\sqrt{16\pi^3}}\frac{m_j}{p^+\sqrt{q^+}},\;\;
V_{2s}(\ub{p},\ub{q})
   \equiv \frac{g}{\sqrt{8\pi^3}}
   \frac{\vec{\epsilon}_{2s}^{\,*}\cdot\vec{p}_\perp}{p^+\sqrt{q^+}},
\ee
with $\vec{\epsilon}_{2s}\equiv-\frac{1}{\sqrt{2}}(2s,i)$. 
The nonzero (anti)commutators are
\bea
\left[a_i(\ub{q}),a_j^\dagger(\ub{q}')\right]
          &=&(-1)^i\delta_{ij}
            \delta(\ub{q}-\ub{q}'), \\
\left\{b_{i,s}(\ub{p}),b_{j,s'}^\dagger(\ub{p}')\right\}
     &=&(-1)^i\delta_{ij}   \delta_{s,s'}
            \delta(\ub{p}-\ub{p}').  \nonumber
\eea

We construct a dressed fermion state, neglecting pair contributions;
it takes the form
\bea
\lefteqn{\Phi_+(\ub{P})=\sum_i z_i b_{i+}^\dagger(\ub{P})|0\rangle
  +\sum_{ijs}\int d\ub{q}_1 f_{ijs}(\ub{q}_1)b_{is}^\dagger(\ub{k})
                                       a_j^\dagger(\ub{k}_1)|0\rangle}&& \\
 && +\sum_{ijks}\int d\ub{q}_1 d\ub{q}_2 f_{ijks}(\ub{q}_1,\ub{q}_2)
       \frac{1}{\sqrt{1+\delta_{jk}}}   b_{is}^\dagger(\ub{k})
                 a_j^\dagger(\ub{k}_1)a_k^\dagger(\ub{k}_2)|0\rangle 
 +\ldots       \nonumber
\eea
The wave functions $f_s(x_n,\vec{q}_{\perp n})$ satisfy the coupled 
system of equations that results from the Hamiltonian eigenvalue
problem $P^+P^-\Phi_+=M^2\Phi_+$.  Each wave function has a total 
$L_z$ eigenvalue of 0 (1) for $s=+1/2$ ($-1/2$).

The coupled equations are
\bea \label{eq:first}
\lefteqn{m_i^2z_i+ \sum_{i',j}(-1)^{i'+j} P^+ \int^{P^+} d\ub{q}
  \left\{ f_{i'j-}(\ub{q})[V_+(\ub{P}-\ub{q},\ub{q})+V_-^*(\ub{P},\ub{q})]
  \right.}&& \\
  &&\left.
  \rule{1in}{0in} 
  + f_{i'j+}(\ub{q})[U_{i'}(\ub{P}-\ub{q},\ub{q})+U_i(\ub{P},\ub{q})]\right\}
   = M^2z_i ,   \nonumber
\eea
\bea \label{eq:oneboson}
\lefteqn{
\left[\frac{m_i^2+q_\perp^2}{1-y}+\frac{\mu_j^2+q_\perp^2}{y}\right]
  f_{ijs}(\ub{q})} &&\\
  &&+\sum_{i'}(-1)^{i'}\left\{
    z_{i'}\delta_{s,-}[V_+^*(\ub{P}-\ub{q},\ub{q})+V_-(\ub{P},\ub{q})] \right. 
    \nonumber \\
  && \rule{1in}{0in}\left.
    +z_{i'}\delta_{s,+}[U_i(\ub{P}-\ub{q},\ub{q})+U_{i'}(\ub{P},\ub{q})]\right\}
    \nonumber \\
    &&+2\sum_{i',k}\frac{(-1)^{i'+k}}{\sqrt{1+\delta_{jk}}}P^+ \int^{P^+-q^+}
    d\ub{q}' \nonumber\\
    && \times\left\{f_{i'jk,-s}(\ub{q},\ub{q}')
       [V_{2s}(\ub{P}-\ub{q}-\ub{q}',\ub{q}')+V_{-2s}^*(\ub{P}-\ub{q},\ub{q}')]
       \right. \nonumber \\
           &&\left.+f_{i'jks}(\ub{q},\ub{q}')
             [U_{i'}(\ub{P}-\ub{q}-\ub{q}',\ub{q}')+U_i(\ub{P}-\ub{q},\ub{q}')]
             \right\} = M^2f_{ijs}(\ub{q}) ,   \nonumber
\eea
and
\bea
\lefteqn{\left[\frac{m_i^2+(\vec{q}_{1\perp}+\vec{q}_{2\perp})^2}{1-y_1-y_2}
    +\frac{\mu_j^2+q_{1\perp}^2}{y_1}+\frac{\mu_k^2+q_{2\perp}^2}{y_2}\right]
        f_{ijks}(\ub{q_1},\ub{q_2})} \\
    &&+\sum_{i'}(-1)^{i'}\frac{\sqrt{1+\delta_{jk}}}{2}P^+ \nonumber \\
    && \times
        \left\{f_{i'j,-s}(\ub{q_1})
[V_{-2s}^*(\ub{P}-\ub{q_1}-\ub{q_2},\ub{q_2})+V_{2s}(\ub{P}-\ub{q_1},\ub{q_2})]
 \right. \nonumber \\
 &&+ f_{i'js}(\ub{q_1})
[U_i(\ub{P}-\ub{q_1}-\ub{q_2},\ub{q_2})+U_{i'}(\ub{P}-\ub{q_1},\ub{q_2})]
 \nonumber \\
 &&  + f_{i'k,-s}(\ub{q_2})
[V_{-2s}^*(\ub{P}-\ub{q_1}-\ub{q_2},\ub{q_1})+V_{2s}(\ub{P}-\ub{q_2},\ub{q_1})]
  \nonumber \\
 &&  \left. + f_{i'ks}(\ub{q_2})
[U_i(\ub{P}-\ub{q_1}-\ub{q_2},\ub{q_1})+U_{i'}(\ub{P}-\ub{q_2},\ub{q_1})]
\right\}+\ldots \nonumber \\
&& \rule{3in}{0in} = M^2f_{ijks}(\ub{q_1},\ub{q_2}). \nonumber
\eea
We consider truncations of this system.

A truncation to one boson leads to an analytically solvable 
problem.\cite{OneBoson}\  The one-boson wave functions are
\bea
f_{ij+}(\ub{q})&=&
   \frac{P^+}{M^2-\frac{m_i^2+q_\perp^2}{1-q^+/P^+}
                -\frac{\mu_j^2+q_\perp^2}{q^+/P^+}}  \\
&\times& \left[\sum_k (-1)^{k+1}z_k)U_i(\ub{P}-\ub{q},\ub{q})
      +\sum_k (-1)^{k+1}z_kU_k(\ub{P},\ub{q})\right],
 \nonumber \\
f_{ij-}(\ub{q})&=&
   \frac{P^+}{M^2-\frac{m_i^2+q_\perp^2}{1-q^+/P^+}
                -\frac{\mu_j^2+q_\perp^2}{q^+/P^+}}
\sum_k (-1)^{k+1}z_k)V_+^*(\ub{P}-\ub{q},\ub{q}).
\eea
Substitution into Eq.~(\ref{eq:first}) yields
\bea \label{eq:onefermion}
(M^2-m_i^2)z_i &=&
 g^2\mu_0^2 (z_0-z_1)J+g^2 m_i(z_0m_0-z_1m_1) I_0
\nonumber \\
  &&+g^2\mu_0[(z_0-z_1)m_i+z_0m_0-z_1m_1] I_1,
\eea
with
\bea
I_n(M^2)&=&\int\frac{dy dq_\perp^2}{16\pi^2}
   \sum_{jk}\frac{(-1)^{j+k}}{M^2-\frac{m_j^2+q_\perp^2}{1-y}
                                   -\frac{\mu_k^2+q_\perp^2}{y}}
   \frac{(m_j/\mu_0)^n}{y(1-y)^n}, \label{eq:In} \\
J(M^2)&=&\int\frac{dy dq_\perp^2}{16\pi^2}
   \sum_{jk}\frac{(-1)^{j+k}}{M^2-\frac{m_j^2+q_\perp^2}{1-y}
                                   -\frac{\mu_k^2+q_\perp^2}{y}}
   \frac{(m_j^2+q_\perp^2)/\mu_0^2}{y(1-y)^2}.
   \label{eq:J}
\eea
The presence of the PV regulators allows $I_0$ and $J$ to satisfy
the identity $\mu_0^2 J(M^2)=M^2 I_0(M^2)$.
With $M$ held fixed, the equations for $z_i$ can be viewed
as an eigenvalue problem for $g^2$.  The solution is
\be \label{eq:gofm}
g^2=-\frac{(M\mp m_0)(M\mp m_1)}{(m_1-m_0)(\mu_0 I_1\pm MI_0)}, \;\;
\frac{z_1}{z_0}=\frac{M \mp m_0}{M \mp m_1}.
\ee
An analysis of this solution is given in Ref.~\refcite{OneBoson}.

In a truncation to two bosons, we obtain the following reduced
equations for the one-boson--one-fermion wave functions:\cite{bhm}
\bea
\lefteqn{\left[M^2
  -\frac{m_i^2+q_\perp^2}{1-y}-\frac{\mu_j^2+q_\perp^2}{y}\right]
f_{ijs}(y,q_\perp)=}&& \\
&&\frac{g^2}{16\pi^2}\sum_a\frac{I_{ija}(y,q_\perp)}{1-y}f_{ajs}(y,q_\perp)
\nonumber \\
 &&  +\frac{g^2}{16\pi^2}\sum_{abs'}\int_0^1dy'dq_\perp^{\prime 2}
   J_{ijs,abs'}^{(0)}(y,q_\perp;y',q'_\perp)f_{abs'}(y',q'_\perp) \nonumber \\
   && +\frac{g^2}{16\pi^2}\sum_{abs'}\int_0^{1-y}dy'dq_\perp^{\prime 2}
   J_{ijs,abs'}^{(2)}(y,q_\perp;y',q'_\perp)f_{abs'}(y',q'_\perp), \nonumber
\eea
where $\sqrt{P^+}f_{ij+}(\ub{q})=f_{ij+}(y,q_\perp)$,
$\sqrt{P^+}f_{ij-}(\ub{q})=f_{ij-}(y,q_\perp)e^{i\phi}$,
$I$ is an analytically computable self-energy,
and $J^{(n)}$ is a kernel determined by $n$-boson intermediate states.
These reduced integral equations are converted to a matrix equation 
via quadrature in $y'$ and $q_\perp^{\prime 2}$.  The matrix
is diagonalized to obtain $g^2$ as an eigenvalue and 
the discrete wave functions from the eigenvector.

A useful set of quadrature schemes is based on Gauss--Legendre 
quadrature and particular variable transformations.
The transformation
for $y'$ is motivated by the need for an accurate approximation
to the integral $J$.  This integral appears implicitly in the
product of the Hamiltonian and the eigenfunction and is
largely determined by contributions near the endpoints whenever the
PV masses are large.  The transformation for the transverse integral 
is chosen to reduce the range from infinite to finite, so that
no momentum cutoff is needed.

\begin{figure}[ht]
\centerline{\includegraphics[width=9cm]{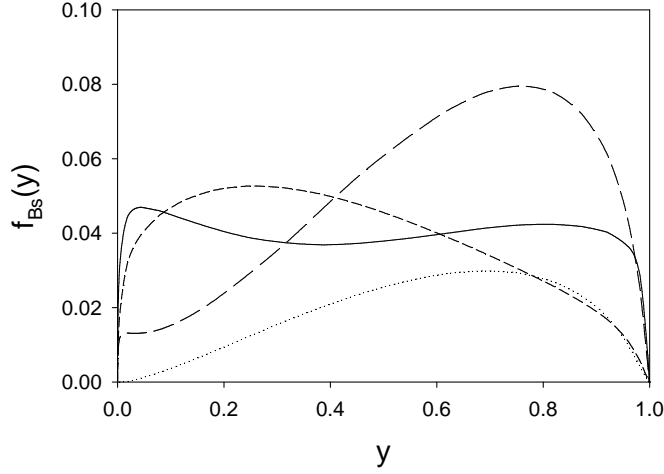}}
\caption{Bosonic structure functions in Yukawa theory,
with a two-boson truncation ($f_{B+}$: solid; $f_{B-}$: long dash)
and a one-boson truncation ($f_{B+}$: short dash; $f_{B-}$: dotted).
The constituent masses had the values $m_0=-1.7\mu_0$, $m_1=\mu_1=15\mu_0$.
The resolutions used in the Gauss--Legendre method are $K=20$ and $N=30$.
} \label{fig:fbs}
\end{figure}

From the wave functions we can extract a structure function $f_{Bs}(y)$,
\bea
\lefteqn{f_{Bs}(y)=\int d\ub{q} \delta(y-q^+/P^+)
   \left|\sum_{ij}(-1)^{i+j}f_{ijs}(\ub{q})\right|^2}&& \\
   &&+\int \prod_{n=1}^2 d\ub{q_n} \sum_{n=1}^2\delta(y-q_n^+/P^+)
   \left|\sum_{ijk}(-1)^{i+j+k}f_{ijks}(\ub{q_n})\right|^2+\ldots,
\eea
defined as the probability density for finding a boson with momentum 
fraction $y$ while the constituent fermion has spin $s$.  Typical 
results are plotted in Fig.~\ref{fig:fbs}.

\section{Feynman-gauge QED}

We apply these same techniques to QED.\cite{qed}  The Feynman-gauge 
Lagrangian is
\bea\label{eq:fgLagrangian}
  {\mathcal L}&=&
    \sum_{i=0}^1 \left(-\frac{1}{4} (-1)^i F_i^{\mu \nu} F_{i,\mu \nu} 
    +(-1)^i \bar{\psi_i} (i \gamma^\mu \partial_\mu - m_i) \psi_i \right. 
    \nonumber \\
&&\left. 
 +  B_i \partial_\mu A_{i}^{\mu} + \frac{1}{2} B_i B_i \right)
 - e \bar{\psi}\gamma^\mu \psi A_\mu,
\eea
where $A^\mu  = \sum_{i=0}^1 A^\mu_i$,
$\psi =\sum_{i=0}^1 \psi_i$, and
$F_i^{\mu \nu} = \partial^\mu A_{i}^{\nu}-\partial_\nu A_{i}^{\mu}$.
The nondynamical fermion fields $\psi_{i-}$ are constrained by
\bea
\lefteqn{i[(-1)^i\partial_-\psi_{i-}-ie\sum_kA_{k-}\sum_j\psi_{j-}]} && \\
  &&=-[(i\gamma^0\gamma^\perp)((-1)^i\partial_\perp\psi_{i+}
     -ie\sum_kA_{k\perp}\sum_j\psi_{j+})-(-1)^im\gamma^0\psi_{i+}]. \nonumber
\eea
For the null combination $\psi_-=\psi_{0-}+\psi_{1-}$, this becomes
\be
i\partial_-\psi_-
    =-[(i\gamma^0\gamma^\perp)\partial_\perp\psi_+-m\gamma^0\psi_+],
\ee
which is independent of $A$ and can therefore be solved without inverting
a covariant derivative.
We then obtain the Hamiltonian without antifermion terms as being
\bea \label{eq:QEDP-}
\lefteqn{P^-=
   \sum_{i,s}\int d\ub{p}
      \frac{m_i^2+p_\perp^2}{p^+}(-1)^i
          b_{i,s}^\dagger(\ub{p}) b_{i,s}(\ub{p})} \\
   && +\sum_{l,\mu}\int d\ub{k}
          \frac{\mu_l^2+k_\perp^2}{k^+}(-1)^l\epsilon^\mu
             a_l^{\mu\dagger}(\ub{k}) a_l^\mu(\ub{k})
          \nonumber \\
   && +\sum_{i,j,l,s,\mu}\int d\ub{p} d\ub{q}\left\{\left[
      b_{i,s}^\dagger(\ub{p}) b_{j,s}(\ub{q})
       V^\mu_{ij,2s}(\ub{p},\ub{q})\right.\right.\nonumber \\
      &&\left.\left.\rule{0.5in}{0in}
+b_{i,s}^\dagger(\ub{p}) b_{j,-s}(\ub{q})
      U^\mu_{ij,-2s}(\ub{p},\ub{q})\right] a_{l\mu}^\dagger(\ub{q}-\ub{p})
                    + h.c.\right\}\,,  \nonumber
\eea
where $\epsilon^\mu = (-1,1,1,1)$.
The vertex functions $U$ and $V$ are given in Ref.~\refcite{qed}.

The dressed electron state, without pair contributions and truncated to
one photon, is
\be
  |\psi\rangle = \sum_i z_i b_{i+}(\ub{P}) |0\rangle + \sum_{s,\mu,i,l}
 \int d\ub{k}f^\mu_{ils}(\ub{k}) b_{is}^\dagger(\ub{k})
               a^\dagger_{l\mu}(\ub{P}-\ub{k}) |0\rangle ,
\ee
with one-photon--one-electron wave functions
\bea
f^\mu_{il+}(\ub{k}) &=& 
    \frac{\sum_j (-1)^j z_j P^+ V_{ij+}^\mu(\ub{k},\ub{P})}
    {1 - \frac{m_i^2 + k_\perp^2 }{ x} 
                  - \frac{\mu_l^2 + k_\perp^2 }{ 1-x}} , \\
f^\mu_{il-}(\ub{k}) &=& 
         \frac{\sum_j (-1)^j z_j P^+ U_{ij+}^\mu(\ub{k},\ub{P})}
     {1 - \frac{m_i^2 + k_\perp^2 }{ x} - \frac{\mu_l^2 + k_\perp^2 }{ 1-x}} .
\eea
Substitution into $P^+P^-|\psi\rangle=M^2|\psi\rangle$ yields
\bea \label{eq:FeynEigen}
(M^2-m_i^2)z_i &=& \frac{\alpha}{2\pi} \int \frac{dx}{x} dk_\perp^2
       \sum_{j,k,l}(-1)^{j+k+l} z_j \\
  && \times    \frac{m_k^2 + k_\perp^2 -2m_k(m_j+m_i)x + m_j m_ix^2}{
      M^2x(1-x) - m_k^2(1-x)  - \mu_l^2x - k_\perp^2}.
\eea
This is the same form as in the one-boson Yukawa problem,
with $g^2\rightarrow 2e^2$ and $I_1\rightarrow -2I_1$, and an
analytic solution is again obtained.  From this solution we can
compute various quantities, including the anomalous magnetic 
moment.\cite{qed}


\section{A correlator in ${\mathcal N}$=(2,2) SYM theory}

Reduction of ${\mathcal N}$=1 SYM 
theory from four to two dimensions
provides the action we need.  In light-cone gauge ($A_-=0$) it is
\bea
\lefteqn{S^{\rm LC}_{1+1}=\int dx^+ dx^- \tr \Bigg[ \del_+ X_I \del_-X_I
 +i\theta^T_R \del^+\theta_R+i\theta^T_L\del^-\theta_L }&& \\
 && +\frac 12 (\del_-A_+)^2+gA_+J^++\sqrt 2 g\theta^T_L\ep_2\beta_I
 [X_I,\theta_R]  +\frac {g^2}4 [X_I,X_J]^2 \Bigg]. \nonumber
\eea
Here the trace is over color indices, the $X_I$ are 
the scalar fields and the remnants of the transverse components of the 
four-dimensional gauge field $A_{\mu}$, the two-component spinors $\theta_R$ 
and $\theta_L$ are remnants of the right-moving and left-moving projections
of the four-component spinor in the four-dimensional theory.  We also
define $J^+=i[X_I,\del_-X_I]+2\theta^T_R\theta_R$,
$\beta_1\equiv\sigma_1$, $\beta_2\equiv\sigma_3$, and $\ep_2\equiv -i\sigma_2$.

The stress-energy correlation function for ${\mathcal N}$=(8,8) SYM theory 
can be calculated on the string-theory side:\cite{N88correlator}
$\bra T^{++}(x)T^{++}(0)\ket=(N_c^{3/2}/g)x^{-5}$. 
We find numerically that this is {\em almost} true in 
${\mathcal N}$=(2,2) SYM theory.\cite{N22}

To compute the correlator,\cite{correlator} we fix the total momentum $P^+$,
compute the Fourier transform, and 
express the transform in a spectral decomposed form
\bea
\tilde F(P^+,x^-)&=&\frac 1{2L}\bra T^{++}(P^+,x^+) T^{++}(-P^+,0)\ket \\
   &=&\sum_i \frac 1{2L}\bra 0|T^{++}(P^+)|i\ket e^{-iP_i^-x^+}
   \bra i|T^{++}(-P^+,0)|0\ket. \nonumber
\eea
The position-space form is recovered by
Fourier transforming with respect to the discrete
momentum $P^+=K\pi/L$, where $K$ is the integer
resolution and $L$ the length scale of DLCQ.\cite{PauliBrodsky}\  This yields
\be  
   F(x^-,x^+)=\sum_i \Big|\frac L{\pi}\bra 0|T^{++}(K)|i\ket\Big|^2\left(
   \frac {x^+}{x^-}\right)^2 \frac{M_i^4}{8\pi^2K^3}K_4(M_i \sqrt{2x^+x^-}) .
   \label{cor}
\ee
We then continue to 
Euclidean space by taking $r=\sqrt{2x^+x^-}$ to be real. 
The matrix element $(L/\pi)\bra 0|T^{++}(K)|i\ket$ is independent of $L$.
Its form can be substituted directly to give an explicit expression for the 
two-point function. 

The correlator behaves like $1/r^4$ at small $r$:
\be 
\left(\frac {x^-}{x^+}\right)^2F(x^-,x^+)
   =\frac{N_c^2(2n_b+n_f)}{4\pi^2r^4}(1-1/K).
\ee
For arbitrary $r$, it can be obtained numerically by either 
computing the entire spectrum (for ``small'' matrices)
or using Lanczos iterations (for large).\cite{correlator}

\begin{figure}[ht]
\centerline{\includegraphics[width=8cm]{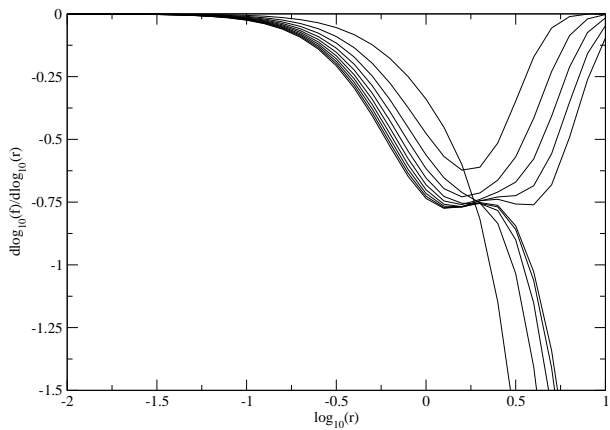}}
\caption{The log derivative of the scaled correlation function $f$
defined in Eq.~(\ref{eq:f}) of the text.  The resolution $K$ ranges 
from 3 to 12.  For even $K$, $f$ becomes constant at large $r$
and the derivative goes to zero.
} \label{fig:Dcor}
\end{figure}

In Fig.~\ref{fig:Dcor}, we plot the log derivative of the scaled 
correlation function\cite{N22}
\be \label{eq:f}
f\equiv \bra T^{++}(x)T^{++}(0)\ket 
      \left(\frac{x^-}{x^+}\right)^2\frac{4\pi^2r^4}{N_c^2(2n_b+n_f)} .
\ee 
At small $r$, the results for $f$ match the expected $(1-1/K)$ 
behavior. At large $r$ the behavior is different between odd and even $K$, 
but as $K$ increases, the differing behavior is pushed to larger $r$. 
For even $K$, there is exactly 
one massless state that contributes to the correlator, while there is no 
massless state for odd $K$. The lowest massive state dominates for odd 
$K$ at large $r$; however, this state becomes massless as $K\to \infty$.
In the intermediate-$r$ region, the correlator behaves like $r^{-4.75}$,
or almost $r^{-5}$. The size of this intermediate region increases
as $K$ gets larger.


\section{${\mathcal N}$=(1,1) SYM theory at finite temperature}

In this case, the Lagrangian is
${\mathcal L}
 ={\rm Tr}\left(-\frac{1}{4}F_{\mu\nu}F^{\mu\nu}
      +i\bar{\Psi}\gamma_{\mu}D^{\mu}\Psi\right)$,
with $F_{\mu\nu}=\partial_{\mu}A_{\nu}-\partial_{\nu}A_{\mu}
+ig[A_{\mu},A_{\nu}]$ and $D_{\mu}=\partial_{\mu}+ig[A_{\mu}]$.
The supercharge in light-cone gauge is
$Q^-=2^{3/4}g\int dx^- \left(i[\phi,\partial_-\phi]
         +2\psi\psi\right)\partial_-^{-1}\psi$.
From the discrete form we can compute the spectrum, which at large-$N_c$
represents a collection of noninteracting modes.  With a sum over
these modes, we can construct the free energy at finite temperature
from the partition function\cite{Elser,FiniteTemp} $e^{-p_0/T}$.

The one-dimensional bosonic free energy is
\be
{\mathcal F}_B=\frac{VT}{\pi}\sum_{n=1}^\infty
\int_{M_n}^\infty dp_0
\frac{p_0}{\sqrt{p_0^2-M_n^2}}
\ln \left( 1- e^{- p_{0}/T }\right),
\ee
and the fermionic free energy is
\be
{\mathcal F}_F= -\frac{VT}{\pi}\sum_{n=1}^\infty
   \int_{M_n}^\infty dp_0
\frac{p_0}{\sqrt{p_0^2-M_n^2}}
 \ln \left( 1+ e^{- p_{0}/T }\right).
\ee
The contributions from the $K-1$ massless states in each sector are
\be
{\mathcal F}^{0}_{B}=-\frac{(K-1)\pi}{6} V T^2, \;\;
{\mathcal F}^{0}_{F}=-\frac{(K-1)\pi}{12} VT^2 .
\ee
The total free energy, with the logs expanded as sums and the
$p_0$ integral already performed, is 
\be
{\mathcal F}(T,V)=-\frac{(K-1)\pi}{4}
   VT^2-\frac{2VT}{\pi}\sum_{n=1}^{\infty}
    {\sum_{l=0}^{\infty}}M_{n}\frac{K_{1}
      \left((2l+1)\frac{M_{n}}{T}\right)}{(2l+1)} .
\ee
The sum over $l$ is well approximated by the first few terms.
We can represent the sum over $n$ as an integral over a density 
of states: $\sum_n \rightarrow \int \rho(M) dM $
and approximate $\rho$ by a continuous function.
The integral over $M$ can then be computed by standard numerical 
techniques.  We obtain $\rho$ by a fit to the computed spectrum 
of the theory and find $\rho(M)\sim \exp(M/T_{\rm{H}})$,
with $T_H \sim 0.845\sqrt{\pi/g^2N_c}$, the Hagedorn 
temperature.\cite{Hagedorn}\  From the free energy we can compute 
various other thermodynamic functions up to this 
temperature.\cite{FiniteTemp}

\section{Future work}

Given the success obtained to date, these techniques are well worth
continued exploration.  In Yukawa theory, we plan to consider the 
two-fermion sector, in order to study true bound states.  For QED
the next step will be inclusion of two-photon states in the calculation
of the anomalous moment.  For SYM theories, we are now able to reach
much higher resolutions, by computing on clusters.  This will permit
continued reexamination of theories where previous calculations were 
hampered by low resolution, particularly in more dimensions.  Earlier 
work on inclusion of fundamental matter\cite{SQCD}\ can be extended 
to three dimensions and modified to include finite-$N_c$ effects, such 
as baryons with a finite number of partons and the mixing of mesons 
and glueballs.  For all of this work, the ultimate goal is, of course, 
the development of techniques sufficient to solve quantum chromodynmics.

\section*{Acknowledgments}
The work reported here was done in collaboration with 
S.J. Brodsky and G. McCartor, and
S. Pinsky, N. Salwen, M. Harada, and Y. Proestos,
and was supported in part by the US Department of Energy
and the Minnesota Supercomputing Institute.


\begin{thebibliography}{0}
%
\bibitem{Dirac} P.A.M. Dirac, 
Rev.\ Mod.\ Phys. \textbf{21}, 392 (1949).
%
\bibitem{PauliBrodsky} H.-C. Pauli and S.J. Brodsky, 
Phys.\ Rev.\ D \textbf{32}, 1993 (1985); 2001 (1985).
%
\bibitem{DLCQreview} S.J. Brodsky, H.-C. Pauli, and S.S. Pinsky, 
Phys.\ Rep.\ \textbf{301}, 299 (1997).
%
\bibitem{lattice} I. Montvay and G. M\"unster,
{\em Quantum Fields on a Lattice}
(Cambridge U. Press, New York, 1994);
J. Smit,
{\em Introduction to Quantum Fields on a Lattice}
(Cambridge U. Press, New York, 2002).
%
\bibitem{PV} W. Pauli and F. Villars,
Rev.\ Mod.\ Phys.\ \textbf{21}, 4334 (1949).
%
\bibitem{bhm-previous} S.J.~Brodsky, J.R.~Hiller, and G.~McCartor,
Phys.\ Rev.\ D \textbf{58}, 025005 (1998) [arXiv:hep-th/9802120];
\textbf{60}, 054506 (1999);
\textbf{64}, 114023 (2001):
Ann.\ Phys.\ \textbf{296}, 406 (2002).
%
\bibitem{OneBoson} S.J. Brodsky, J.R. Hiller, and G. McCartor,
Ann.\ Phys.\ \textbf{305}, 266 (2003).
%
\bibitem{bhm} S.J. Brodsky, J.R. Hiller, and G. McCartor, in preparation.
%
\bibitem{qed} S.J. Brodsky, V.A. Franke, J.R. Hiller, G. McCartor,
S.A. Paston, and E.V. Prokhvatilov, arXiv:hep-ph/0406325.
%
\bibitem{SDLCQreview} O.~Lunin and S.~Pinsky,
   in {\em New Directions in Quantum Chromodynamics}, 
   edited by C.-R.~Ji and D.-P.~Min,
   AIP Conf.\ Proc.\ No.\ 494 (AIP, Melville, NY, 1999), p.~140,
   [arXiv:hep-th/9910222].
%
\bibitem{3Dsdlcq}
F.~Antonuccio, O.~Lunin, and S.~Pinsky,
Phys.\ Rev.\ D {\bf 59}, 085001 (1999);
P.~Haney, J.R.~Hiller, O.~Lunin, S.~Pinsky, and U.~Trittmann,
Phys.\ Rev.\ D {\bf 62}, 075002 (2000);
J.R.~Hiller, S.~Pinsky, and U.~Trittmann,
Phys.\ Rev.\ D {\bf 64}, 105027 (2001).
%
\bibitem{N88correlator}
F.~Antonuccio, A.~Hashimoto, O.~Lunin, and S.~Pinsky,
JHEP {\bf 9907}, 029 (1999).
%
\bibitem{correlator}
J.R.~Hiller, O.~Lunin, S.~Pinsky, and U.~Trittmann,
Phys.\ Lett.\ B {\bf 482}, 409 (2000);
J.R.~Hiller, S.~Pinsky, and U.~Trittmann,
Phys.\ Rev.\ D {\bf 63}, 105017 (2001). 
%
\bibitem{N22} M. Harada, J.R. Hiller, S. Pinsky, and N. Salwen,
to appear in Phys.\ Rev. D,
arXiv:hep-ph/0404123.
%
\bibitem{FiniteTemp} J.R. Hiller, Y. Proestos, S. Pinsky, and N. Salwen,
arXiv:hep-th/0407076.
%
\bibitem{Sakai} Y.~Matsumura, N.~Sakai, and T.~Sakai,
Phys.\ Rev.\ D {\bf 52}, 2446 (1995).
%
\bibitem{Elser}
S.~Elser and A.C.~Kalloniatis,
Phys.\ Lett.\ B {\bf 375}, 285 (1996).
%
\bibitem{Hagedorn}
R. Hagedorn, Nuovo Cimento Suppl.\ {\bf 3}, 147 (1965); 
Nuovo Cimento {\bf 56A}, 1027 (1968).
%
\bibitem{SQCD} J.R. Hiller, S.S. Pinsky, and U. Trittmann,
Nucl.\ Phys.\ B {\bf 661}, 99 (2003);
Phys.\ Rev.\ D {\bf 67}, 115005 (2003).
%
\end{thebibliography}
\end{document}